%% file: main.tex
\documentclass[
    aip,
    reprint,
    amsmath,amssymb,
    ]{revtex4-1}

\usepackage[utf8]{inputenc}
\usepackage{hyperref}
\usepackage{color}
\usepackage{listings}
\usepackage[frozencache,cachedir=minted-cache]{minted}
\usepackage{graphicx}
\usepackage{dcolumn}
\usepackage{bm}
\usepackage{rotating}
\usepackage{afterpage}
\usepackage[table,xcdraw]{xcolor}
\usepackage[normalem]{ulem}
\usepackage[utf8]{inputenc}
\usepackage[T1]{fontenc}
\usepackage{mathptmx}
\usepackage{etoolbox}
\useunder{\uline}{\ul}{}

\setminted[python]{
    fontsize=\footnotesize,
}

\makeatletter
\def\@email#1#2{%
 \endgroup
 \patchcmd{\titleblock@produce}
  {\frontmatter@RRAPformat}
  {\frontmatter@RRAPformat{\produce@RRAP{*#1\href{mailto:#2}{#2}}}\frontmatter@RRAPformat}
  {}{}
}%
\makeatother

\graphicspath{{figures/}}

\newcommand{\ch}[1]{{#1}}
\newcommand{\markuponly}[1]{{}}

\newcommand{\py}[1]{{\tt #1}}
\newcommand{\wfl}{\py{wfl} }

\input{insbox}

\begin{document}

\title{\py{wfl} Python Toolkit for Creating Machine Learning Interatomic Potentials and Related Atomistic Simulation Workflows}
\author{Elena Gelžinytė*}
    \affiliation{Engineering Laboratory, University of Cambridge, Trumpington Street, Cambridge CB2 1PZ, United Kingdom}
    \email{eg475@cam.ac.uk}
\author{Simon Wengert}
    \affiliation{Fritz-Haber-Institut der Max-Planck-Gesellschaft, Faradayweg 4-6, D-14195 Berlin, Germany}
\author{Tam\'as K. Stenczel}
    \affiliation{Engineering Laboratory, University of Cambridge, Trumpington Street, Cambridge CB2 1PZ, United Kingdom}
\author{Hendrik H. Heenen}
    \affiliation{Fritz-Haber-Institut der Max-Planck-Gesellschaft, Faradayweg 4-6, D-14195 Berlin, Germany}
\author{Karsten Reuter}
    \affiliation{Fritz-Haber-Institut der Max-Planck-Gesellschaft, Faradayweg 4-6, D-14195 Berlin, Germany}
\author{G\'abor Cs\'anyi}
    \affiliation{Engineering Laboratory, University of Cambridge, Trumpington Street, Cambridge CB2 1PZ, United Kingdom}
\author{Noam Bernstein}
    \affiliation{Center for Materials Physics and Technology, U.~S. Naval Research Laboratory Code 6393, 4555 Overlook Ave SW, Washington, DC 20375, United States of America}
\date{\today}

\begin{abstract}
\InsertBoxL{0}{\includegraphics[width=0.15\linewidth]{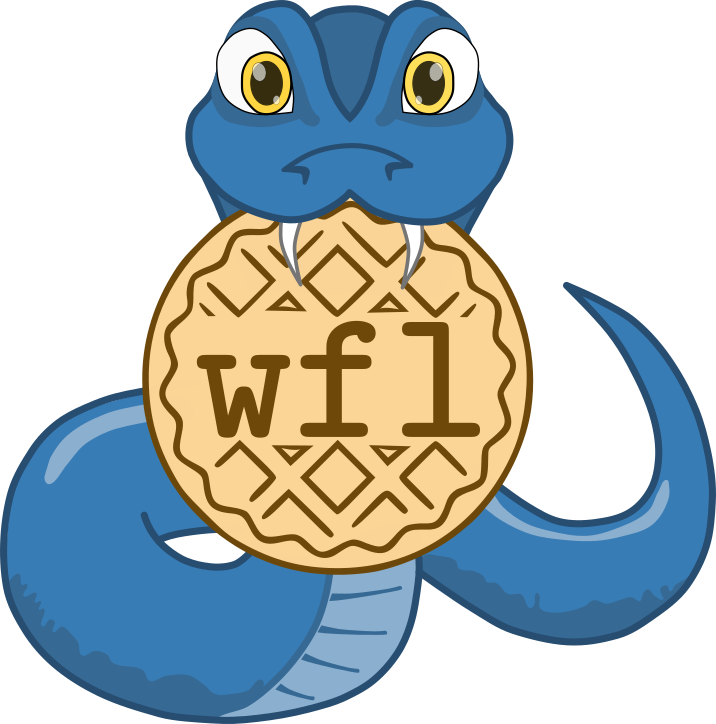}}
\noindent Predictive atomistic simulations are increasingly employed for data intensive high throughput studies that take advantage of constantly growing computational resources. To handle the sheer number of individual calculations that are needed in such studies, workflow management packages for atomistic simulations have been developed for a rapidly growing user base. These packages are predominantly designed to handle computationally heavy \emph{ab initio} calculations, usually with a focus on data provenance and reproducibility. However, in related simulation communities, e.g.\ the developers of machine learning interatomic potentials (MLIPs), the computational requirements are somewhat different: the types, sizes, and numbers of computational tasks are more diverse, and therefore require additional ways of parallelization and local or remote execution for optimal efficiency.
In this work, we present the atomistic simulation and MLIP fitting workflow management package \py{wfl} and Python remote execution package \py{ExPyRe} to meet these requirements. 
With \py{wfl} and \py{ExPyRe}, versatile Atomic Simulation Environment
based workflows that perform diverse procedures can be written. This capability is based on a low-level developer-oriented framework, which can be utilized to construct high level functionality for user-friendly programs. Such high level capabilities to automate machine learning interatomic potential fitting procedures are already incorporated in \py{wfl}, which we use to showcase its capabilities in this work. We believe that \wfl fills an important niche in several growing simulation communities and will aid the development of efficient custom computational tasks. 

\end{abstract}

\maketitle

\input{introduction.tex}

\input{examples.tex}

\input{examples_writing.tex}

\input{outro.tex}

\bibliographystyle{plain} 
\bibliography{refs} 

\end{document}

%% file: introduction.tex
\section{Introduction}
\label{sec:introduction}

It is common to perform a large number of expensive calculations in computational chemistry and materials science. 
In many cases these tasks are embarrassingly parallel, i.e.\ each calculation can be performed completely independently of the rest.
Some examples include single point calculations, geometry optimisation, spectra and similar prediction of large databases of atomistic structures, high-throughput screening (e.g.\ random structure search), or generating reference data for fitting machine-learning or conventional interatomic potentials. 
Due to the throughput enabled by modern High Performance Computing (HPC) it is no longer practical for these calculations be launched and monitored manually, and a number of packages to manage such workflows have recently been developed. 

Most of the workflow packages focus on building \textit{ab initio} material property databases \cite{gjerding2021atomic,mathew2017atomate,jain2015fireworks, pizzi2016aiida,janssen2019pyiron, kirklin2015open,choudhary2020joint}. 
The frameworks define workflows to calculate certain structural and electronic properties, for example band structures, spectra or dielectric constants, and are themselves made up of modular steps, e.g.\ geometry optimisation, structure perturbations, and single point calculations. These packages provide a consistent way to build and extend the large databases, e.g.\ Materials Project \cite{jain2013commentary}, Open Quantum Materials Database \cite{saal2013materials}, Aflow \cite{esters2023aflow}, and pay particular attention to reproducibility and data provenance. To go in hand with large projects, most of the existing packages tend to provide a relatively structured and user-focused interface for (re-)running the workflows, analysing and querying the results. 
Such packages need to efficiently carry out the same workflow over many structures, but may also execute related \textit{ab initio} workflows for a given structure, for example, to screen different DFT settings for convergence or calculate different electronic properties for a given ground state. 
Table \ref{table:other_packages} compares some of the popular atomistic workflow managing packages, paying particular attention to the points relevant to our applications, as discussed further in Section \ref{sec:introduction:requirements}.

\begin{table*}[]
\begin{tabular}{l p{1.6cm} l p{2.1cm} p{3.6cm} p{1.8cm} p{3.0cm} }
\hline
Package & Atomistic Structures & Data on Disk     & Autoparallelize over configs & Queued Execution & Interface & Interface with\newline Remote Cluster \\ \hline \hline
\py{wfl}     & ASE\cite{larsen2017atomic}                                                            & File (\py{ase.io}) & Yes                                                                   & Yes (ExPyRe)                                                & Python  & SSH                                                                     \\ \hline
ASR\cite{gjerding2021atomic}     & ASE\cite{larsen2017atomic}                                                            & ASE DB, File        & -                                                                    & Local only (MyQueue\cite{mortensen2020myqueue})                                                 & CLI, Python  & -                                                                       \\ \hline
Atomate\cite{mathew2017atomate} & Pymatgen\cite{ong2013python}                                                       & MongoDB          & -                                                              & Yes (FireWorks\cite{jain2015fireworks})                                             & CLI, Python    & Network to central DB                                                   \\ \hline
AiiDA\cite{pizzi2016aiida}   & Own                                                            & SQL              & -                                                                    & Yes                                                         & CLI, Python   & Daemon on cluster             \\ \hline
PyIron\cite{janssen2019pyiron}  & Own                                                            & File, SQL, HDF5     & -                                                                    & Yes (Pysqa)                                                 & Python        & SSH                                                                     \\ \hline
Aflow\cite{curtarolo2012aflow}   & Own                                                            & File                & -                                                                    & -                                                         & CLI & -                                                                       \\ \hline
Icolos\cite{moore2022icolos}  & RDKit\cite{rdkit}                                                          & File                & -                                                                   & Local only                                                          & CLI  & -                                                                       \\ \hline
qmpy\cite{kirklin2015open}    & Own                                                            & File, MySQL         & -                                                                    & Yes                                                        & CLI, Python   & SSH                                                                     \\ \hline
JARVIS\cite{choudhary2020joint}  & Own                                                            & File                & -                                                                    & Local only                                                          & Python        & -                                                                       \\ \hline
\end{tabular}
\caption{Key features and dependencies of packages for automated atomistic simulations workflows. ``CLI'' indicates
command-line interface.
``File'' indicates format specific to package unless otherwise specified in parentheses.
``Autoparallelize over configs'' refers to the capacity to readily parallelize custom \emph{operations} with a wide range of computational requirements (microseconds-days) and over a wide range \ch{in the number} of atomic structures (10s-100,000s).
}
\label{table:other_packages}
\end{table*}


Machine-learning interatomic potentials (MLIPs) have recently flourished, also enabled by the large amounts of data producible by modern HPC resources. These models are fitted to electronic structure data to approximate the reference potential energy surface (PES) at a much lower computational cost. In addition to relatively few (100s-1000s), but computationally expensive (hours-months) reference data evaluations, such workflows also require considerably cheaper (micro-miliseconds), but substantially more numerous (10~000s-100~000s) energy and force evaluations, which is a mode of operation not targeted by the currently existing packages. 
Filling this niche, \ch{in this work} we introduce the \wfl \ch{and \py{ExPyRe}} package\ch{s}.

The aim of the \wfl package is to support high-throughput execution of the wide variety of tasks encountered in MLIP fitting and atomistic simulation workflows. 
\ch{The \py{ExPyRe} package handles the (remotely) queued execution of Python functions and while used extensively in \py{wfl}, it is independent and can be applied to any Python function.}
In our projects, we rely extensively on the broad range of tools available through the Atomic Simulation Environment~\cite{larsen2017atomic} (ASE) and thus have built \wfl as a lightweight extension to ASE-based scripts with low barrier of entry for those already used to developing atomistic simulations with Python and ASE. The focus of \wfl is less on reproducibility and more on flexibility and ease of development. The main tools, namely input/output abstraction, autoparallelization, and remote execution, are lower-level than those in most of the existing packages and are more developer- rather than user-focused. To facilitate this, we select formats that allow workflow steps to be human-inspectable, and do not rely on client-server databases for ease of operation on restrictive centrally managed HPC clusters. While \wfl includes a substantial number of modular functions already wrapped in \wfl functionality, we emphasise that \ch{any MLIP fitting framework with a Python or command line interface can be integrated into} \py{wfl}-based workflows\ch{. They are also} easily extendable with new \emph{operations}, which can then automatically gain \py{wfl}'s autoparallelisation and\ch{, via \py{ExPyRe},} remote execution functionalities.

In Section \ref{sec:introduction:requirements} we describe key principles guiding the design of \py{wfl} and give a general outline of the main types of tools provided by \py{wfl}. Section \ref{sec:using_by_example} contains a number of code snippet examples for using various higher-level functions already implemented in \py{wfl}. Finally, Section \ref{sec:writing_functions} shows how to 
implement new operations by wrapping functions using the low-level \py{wfl} functionality. The online documentation (\url{libatoms.github.io/workflow}) and docstring-based documentation 
have more complete examples and include an exhaustive list of available functionalities and optional arguments.

\section{Overview}
\label{sec:introduction:requirements}

\subsection{Design principles}

Machine-learning interatomic potentials  are increasingly widely used in atomistic simulations, providing \textit{ab initio} accuracy at dramatically lower computational cost. One of the components in building MLIPs is high-throughput \textit{ab initio} evaluations to collect reference data, a task that is catered to by most of the existing atomistic workflow packages. \ch{Indeed, a number of such packages have been used to build \textit{ab initio} databases for fitting MLIPs~\cite{poul2023systematic,dragoni2018achieving,marchand2020machine,kloppenburg2023general,mirhosseini2021automated}.}
However, other tasks, while also embarrassingly parallel and/or computationally expensive, are particular to MLIP-building workflows and require somewhat different considerations, \ch{not yet generally catered to by the existing packages}. These include, but are not limited to, fitting the MLIP and using it to drive simulations which yield new atomic structures that need to be sub-selected for further processing. Likewise, atomistic simulation workflows that {\em use} MLIPs operate in a somewhat different throughput regime due to much lower computational cost of MLIPs compared to electronic structure codes.

In the case of \textit{ab initio} reference data collection,
each evaluation is expensive enough to take up anywhere from one to potentially hundreds of nodes of an HPC, and from minutes to hours of wall time. At the other end of the scale is the evaluation of significantly computationally cheaper operations over many orders of magnitude more configurations (e.g.\ evaluate the fitted interatomic potential, or calculate a descriptor), where each calculation may take only a fraction of a second on a single CPU core. The third type is ``one-off'' expensive tasks, such as fitting an MLIP or sub-selecting the most diverse structures from the database with potentially complex algorithms, e.g.\ CUR decomposition. In many cases the codes that carry out these tasks are themselves parallelized, but have high memory requirements or are run on GPUs, for example. Finally, there  are often other ASE \py{Atoms}-based project-specific \textit{ad hoc} tasks that need parallelization or remote execution, which should be easy to add to the workflow management package.

From a practical point of view of the developer, code often grows organically over the life of a project, so modularity is of essence when adding functionality as the project develops. Further, it is helpful if the workflow is easy to prototype and then also easy to scale-up from run to run, for example to develop and quickly test the code locally on a small database with a cheap stand-in reference method and easily change to large databases, expensive electronic structure codes and execution on a HPC cluster. When running, it is desirable to have enough control over each of the resources for each modular part---for example, submitting expensive MLIP fitting to a cluster, but evaluating the cheap potentials and analysing the results locally. 
\ch{
Finally, once the code has stabilized and the workflow is applied to a production problem, the computational time is often long, 
and it is helpful if the process can be restarted after interruptions without repeating all of the already completed computations. 
}

\subsection{Technical requirements}

Alongside the modular, developer-oriented package design principles, we had a number of technical requirements. First, we choose compatibility with ASE, which most of our simulation workflows were already based on. ASE is widely used for atomistic simulations and supports a wide range of tasks (from equation of motion integrators to building atomic structures) and has a unified interface to many electronic structure (first principles and semi-empirical) codes and force field libraries. Some of the current atomistic workflow packages (see Table \ref{table:other_packages}) are based on ASE~\cite{larsen2017atomic}, Pymatgen~\cite{ong2013python} and RDKit~\cite{rdkit}, but a considerable number of them implement custom data structures and interfaces with electronic structure codes.

Secondly, the package was not to rely on client/server-based databases or daemons, because HPC clusters often prohibit setting up unmanaged background processes or opening network ports. This requirement is in contrast to how many of the other workflow frameworks manage the large number of calculations running on the HPC clusters.

Third, the data (simulation results, notes on executed code, etc.) are often similarly stored in databases, whereas we prefered human-readable files to facilitate manual inspection of atomic structures, debugging, and error-detection.   
In principle any \py{ase.io} format may be used in \py{wfl}, but in practice \py{extxyz} is recommended because the package assumes that arbitrary per-configuration (\py{Atoms.info}) and per-atom (\py{Atoms.arrays}) quantities for each \py{Atoms} object are stored.

Finally, \wfl needed to have a low barrier to entry for simulation workflow developers familiar with Python and ASE. 
Therefore, the provided extensions are designed to be modular and have a lightweight interface, yet still maintain a good amount of flexibility. As a result, it is straightforward to modify ASE-only code to take advantage of \wfl functionality.

\subsection{Key tools \py{wfl} provides}

The primary tools \wfl provides are low-level functions for extending ASE-based scripts. 
A core concept in \wfl is an \textit{operation}---a Python function that acts on or creates a number of atomic structures. Most of the ASE functionalities are focused on handling one structure (i.e.\ \py{Atoms} object) at a time, whereas with \wfl operations can be parallelized over \py{Atoms} and/or executed remotely and mixed-and-matched to create a complex MLIP fitting, simulation, or analysis workflow. That is because the key utilities of \wfl are practically entirely generic. 
The paralellization functionality is function-agnostic: any operation on a single \py{Atoms} object, for example evaluation with any ASE-supported \py{Calculator}, can be parallelized over a set of them. In a similar way, virtually any Python function can be executed remotely by ExPyRe - a separate package for submitting Python functions as jobs to a cluster's queueing system. The tools together are made for it to be straightforward to use \wfl for small tasks or to prototype simulations and then scale them up to complex and task-heterogeneous workflows. 
See Section \ref{sec:using_by_example} for examples that use the abstracted input/output classes and autoparallelized functions and Section \ref{sec:writing_functions} for detailed examples of how to add \py{wfl} functionality to existing ASE-based operations and scripts.

While the low level \py{wfl} components (file I/O, autoparallelization, remote execution) are designed to be easily integrated with ASE-based Python scripts, \py{wfl} also covers a number of higher-level operations that already take advantage of these capabilities.
First, applying any ASE \py{Calculator}  to a set of \py{Atoms} objects can be parallelized via \py{wfl.calculators.generic.calculate}. This covers a wide variety of codes supported by ASE and ML potentials which often have an interface via ASE's \py{Calculator} class. A special case are electronic structure codes, which are almost exclusively file-based.
Therefore, the corresponding \py{wfl} wrappers have to be modified so that parallel instances do not interfere with each other. 
Thus, \py{wfl}-parallelization-compatible derived classes are included for ORCA, CASTEP, VASP,  Quantum ESPRESSO and FHI-Aims codes. 

Additional common per-\py{Atoms}-parallelizable operations are also included. Structures can be generated from SMILES and AIRSS \py{buildcell} \cite{Pickard2011airss},
perturbed by normal-mode or phonon displacements, selected with simple or complex criteria, and evolved with molecular dynamics, global or local geometry optimisation. Other operations that are not embarrassingly parallel include the 
selection of structures via farthest point sampling or CUR decomposition on atomic descriptors, as well as interfaces with Gaussian Approximation Potential (GAP)~\cite{bartok2010gaussian} and Atomic Cluster Expansion (ACE)~\cite{drautz2019atomic,dusson2022atomic} fitting codes for the full data-fit-repeat cycle. 

Together these features provide a versatile collection of modular tools to mix-and-match for building interatomic potentials beginning-to-end and running complex atomistic simulations. The principal focus of \py{wfl} is on making ASE and Python-based atomistic simulation scripts easier to develop and scale up by parallelization and remote execution.

\textbf{}

%% file: examples.tex
\section{Using \py{wfl} by Example}
\label{sec:using_by_example}

We illustrate the concepts that motivate the design of \py{wfl} through a
series of simple examples that demonstrate their usage. We begin with the basic
Python classes that facilitate operations on a sequence of configurations,
\py{ConfigSet} and \py{OutputSpec}, and controlling the parallelization of this computation on a single 
computer using the \py{AutoparaInfo} class. We then use an example of a more computationally
demanding calculation to show how the work can be further parallelized over multiple
nodes by submitting independent jobs to a queuing system with \py{ExPyRe} and the
\py{wfl} \py{RemoteInfo} class. We also include an example of a command to fit a GAP potential - a computationally expensive task that
is dispatched by \py{wfl}, but parallelized in the \py{gap\_fit} code itself. Finally, we highlight how multiple operations can be daisy-chained into a workflow by passing outputs of one operation as inputs to another and briefly describe a more complex workflow given as an example in the SI and online documentation\footnote{\url{https://libatoms.github.io/workflow/examples.daisy_chain_mlip_fitting.html}}.

\subsection{Atomic configuration input and output abstraction}
\label{sec:using:io_abstraction}

The basic operation that \py{wfl} is designed to facilitate is an embarrassingly parallel
application of the same task on each of a large number of configurations, resulting
in a set of output configurations that maps one-to-one with the input set. The input for
such an operation is specified using the \py{ConfigSet} class. The configurations
can be stored in one file
\begin{minted}{python}
inputs = ConfigSet("configs.xyz")
\end{minted}
multiple files
\begin{minted}{python}
in_files = list(glob.glob("configs/c*.xyz"))
inputs = ConfigSet(in_files)
\end{minted}
or a pre-existing list of configurations in memory, here created on-the-fly via ASE
\begin{minted}{python}
in_configs = [Atoms(numbers=[i], cell=[10]*3, pbc=[True]*3)
              for i in range(1, 10)]
inputs = ConfigSet(in_configs)
\end{minted}
For files, since \py{wfl} uses \py{ase.io.read} to read the configurations, any compatible file 
format is allowed, and optional arguments can be passed in \py{read\_kwargs}.  The output of 
the operation is specified using the \py{OutputSpec} class. If the output configurations only 
needs to be saved in memory, without being backed up by persistent file storage, the 
constructor can be called without arguments
\begin{minted}{python}
outputs = OutputSpec()
\end{minted}
Passing one or more filenames results in the output configurations being stored to
file-based storage, which is non-volatile and therefore available for inspection by the
user or for a restart of the workflow. The basic syntax is identical to that of \py{ConfigSet}:
\begin{minted}{python}
outputs = OutputSpec("evaluated_configs.xyz")
\end{minted}
or
\begin{minted}{python}
output_files = [Path(f).parent / ("calc" + Path(f).name) 
                for f in glob.glob("configs/c*.xyz")]
outputs = OutputSpec(output_files)
\end{minted}
Note that the output can always be written to a single file, but because of the one-to-one mapping, 
if multiple {\em output} files are specified, the number must match the number of input files.

This range of possible input and output targets makes it possible to write workflows that are
relatively independent of how the initial, intermediate, and final configurations are stored.
For any combination, the operation that performs the embarrassingly parallel operation on
each configuration and returns a resulting configuration is called with the same syntax.
For example, if the energy with \py{ASE}'s effective medium theory (EMT) is needed, the most
basic call would use

\begin{minted}{python}
from wfl.calculators import generic
evaluated_configs = generic.calculate(
    inputs, outputs, calc=EMT(),
    property_prefix="evaluated_")
\end{minted}

\py{ASE} stores calculated properties in the \py{Atoms.calc} \py{Calculator} object, which
has two shortcomings for the uses we envision.  The first is that the \py{Calculator} is not 
preserved when \py{Atoms} is written to a file.  The second is that only one calculator,
and hence one set of results, can be associated with an \py{Atoms} object, so it is not possible
to keep, e.g., both DFT reference and tested potential results. Therefore, the \py{wfl} wrapper \py{generic.calculate}
saves the properties in the resulting configurations' \py{Atoms.info} (per-config) and \py{Atoms.arrays} (per-atom) dictionaries.
The keys include a prefix, to distinguish evaluations with different calculators, which
defaults to the name of the calculator class, but in this example is overridden by 
the optional \py{property\_prefix} argument \py{"evaluated\_"}, so the keys will be \py{evaluated\_energy}, etc.
The returned value of the \py{evaluated\_configs} variable is a \py{ConfigSet} object pointing
to the resulting configurations, regardless of whether the \py{outputs} \py{OutputSpec}
indicated memory, single file, or multiple file storage. This new \py{ConfigSet} can then be 
passed as the inputs to the next step in the workflow, with a new \py{OutputSpec} indicating 
where the next step's results will be stored (see Section \ref{sec:using:daisy_chain}).

\begin{figure*}[ht]
    \centering
    \includegraphics[width=\linewidth]{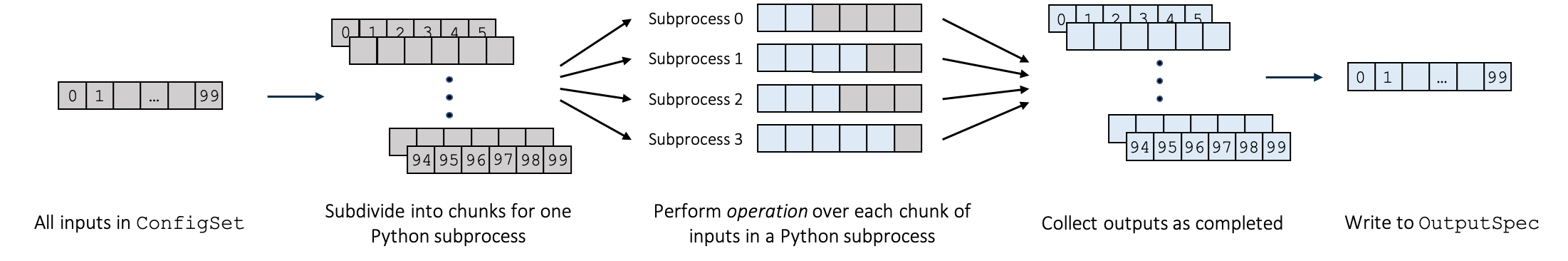}
    \caption{Illustration of the autoparallelization mechanism. Colors indicate the status of each configuration:
    gray for unprocessed, and blue for a completed operation. 
    }
    \label{fig:simple_parallelization}
\end{figure*} 

One advantage to storing intermediate steps in files is that they can be used to recover
from a partially complete sequence of operations. To facilitate this possibility, the default behavior for \py{wfl} 
operations is to {\em skip over the actual operation} if \py{OutputSpec} indicates file storage {\em and} all the 
files already exist. To avoid mistaking partially complete operations for successfully completed ones, the low 
level routines initially write the output to temporary filenames, and only rename them to the intended final 
names once the operation is complete. 
\ch{Note that this mechanism is very simple, relying only on the existence of files with particular names, so
it is not aware of changes in the {\em code}. As a result, during development it is necessary for the user to manually delete
any output files created by functions that have been modified since the previous run.}
\ch{Also, writing the intermediate results to file storage does lead to a small additional computational cost. For example, the simple workflow example discussed in Section~\ref{sec:using:simple_workflow} is about 4\% (23~s) slower if all of the \py{OutputSpec}s write to an extended xyz file instead of keeping the structures in-memory.\footnote{The comparison is made on the example workflow script, appropriate steps parallelized over 16 cores with no remote execution, and 10 times more starting SMILES strings and structures in the training and testing sets. }}

\subsection{Simple autoparallelization}
\label{sec:using:simple_autopara}

Since this interface is designed for operations that can be done independently for each
configuration, it is trivial to parallelize, as illustrated in Fig.~\ref{fig:simple_parallelization}.
The implementation has been encapsulated by
a Python wrapper defined in \py{wfl.autoparallelize},
but a full description of its details are beyond the 
scope of this article.  Using this automatic parallelization is very simple, and can be controlled by
a combination of an \py{AutoparaInfo} object and an environment variable, both optional.
In the simplest case, the Python calling syntax is
as shown above, and the only additional requirement for activating the parallelization is to set the environment variable  \py{WFL\_NUM\_PYTHON\_SUBPROCESSES="N"}, 
where \py{N} is the number of Python processes to parallelize over. 
It is also possible to set this 
value from Python, by passing an \py{AutoparaInfo} object
\begin{minted}{python}
from wfl.autoparallelize import AutoparaInfo
autopara_info = AutoparaInfo(num_python_subprocesses=8)
evaluated_configs = generic.calculate(
    inputs, outputs, calc=calc, 
    autopara_info=autopara_info)
\end{minted}
The \py{autoparallelize} wrapper ensures that all autoparallelized functions  take \py{inputs} and \py{outputs} 
as their first two arguments, and the \py{autopara\_info} keyword argument. The sequence of configurations given as inputs are broken up into chunks and passed to a number of Python processes (defined by the environment variable or \py{AutoparaInfo} argument) which
execute the low-level function implementing the operation. Returned configurations are reassembled according to the original \py{inputs} sequence and written to the location indicated by \py{outputs}.

One limitation of this design is that all arguments passed to and results returned from the function
must be picklable, because Python's \py{multiprocessing.pool}, which we use, depends on the \py{pickle} module 
to communicate with the subprocess running the called function. One important use case, GAP ASE calculator class \py{Potential} defined in \py{quippy.potential},
does not satisfy this requirement.  As 
a result, the \py{generic.calculate} wrapper will also accept a 3-element tuple, instead of an instantiated \py{Calculator} 
object, with the structure
\begin{minted}{python}
calc = (quippy.potential.Potential, 
        [], {"param_filename"="GAP.xml"})
\end{minted}
where the first element is the constructor method, the second is a list of positional arguments, and the
third is a dictionary of keyword arguments.

\subsection{Parallelizing expensive operations over independent queued jobs}
\label{sec:using:remote_jobs}

\begin{figure*}[ht]
    \centering
    \includegraphics[width=\linewidth]{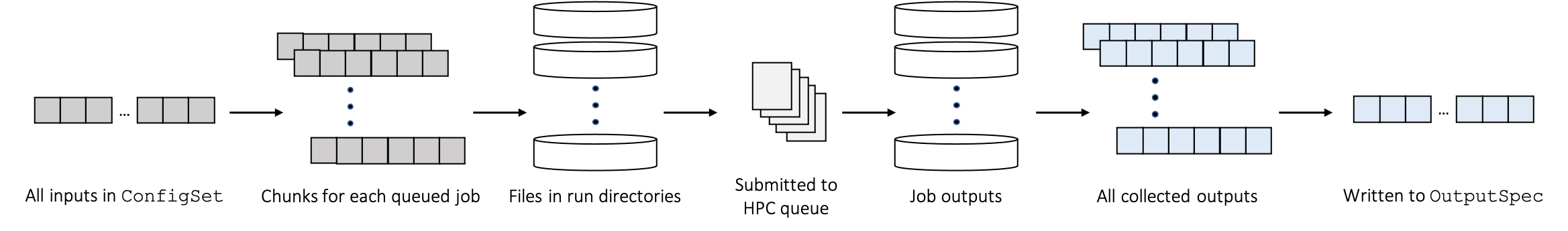}
    \caption{Illustration of autoparallelization mechanism with remote execution (colors as in Fig.~\ref{fig:simple_parallelization}). Within each queued job further parallelization can be carried out, as described in Section~\ref{sec:using:simple_autopara} and Figure~\ref{fig:simple_parallelization}.}
    \label{fig:autopara_remote}
\end{figure*}

In addition to facilitating parallelization over Python subprocesses, the \py{autoparallelize} wrapper 
also makes it easy to break up the work into a set of independent jobs for execution with a queuing system, illustrated in 
Fig.~\ref{fig:autopara_remote}. This capability is especially important for operations where the cost of
application to a single configuration is substantial, e.g.\ single-point DFT evaluations of a potential
fitting database. This functionality is provided using the new \py{ExPyRe} Python package for {\tt Ex}ecuting {\tt Py}thon {\tt Re}motely. 
From the point of view of the \py{wfl} user, the only additional requirement is to specify information about the remote
job in the \py{AutoparaInfo} argument, including the computer system where it will run and the resources required.
\py{ExPyRe} is designed for HPC facilities with a queuing system\ch{---PBS, SLURM and SGE are currently supported. 
T}he computer where the script
using \py{wfl} runs (but not necessary the HPC system where the jobs will run) must have a configuration file describing the available systems and resources where the queued jobs 
can be submitted, as well as the \py{wfl} and \py{ExPyRe} Python packages.  This workflow-running machine can be a
login node of the HPC system, or a different computer whose job submission node is accessible via ssh.
The HPC system compute nodes must have Python and \py{wfl} installed, but no other Python packages are required
except what is necessary to carry out the actual desired computation, e.g.\ \py{quippy} or \py{phonopy}.

The configuration information is by default stored at \py{"$\sim$/.expyre/config.json"}, and contains a \py{dict}  with a \py{systems} key with entries such as 

\begin{samepage}
\begin{minted}[fontsize=\footnotesize]{json}
{"systems": {
  "local_HPC": {
    "host": null,
    "scheduler": "sge",
    "commands": [
      "conda activate myenv"
    ],
    "header": [
      "#$ -pe smp {num_cores}"
    ],
    "partitions": {
      "standard": {
        "num_cores": 16,
        "max_time": "168h",
        "max_mem": "200GB"
}}}}}
\end{minted}
\end{samepage}

This example describes a system with no remote host (i.e.\ jobs will be submitted on the machine where the 
\py{wfl}-using script runs), using the SGE queuing system with one optional queued job header line, and a 
single \py{conda} command to run at the start of any job. The system has only one partition type, named ``standard'', 
which has the specified node core count, memory, and time limit. \py{ExPyRe} will use this
information to create the job script, including the mandatory header lines, which will be submitted
for each remote job.

With this configuration in place, breaking up the \py{inputs} into groups to be run in a series
of queued jobs requires only a small addition to the \py{generic.calculate} calling syntax, an optional
argument to the \py{AutoparaInfo} constructor: 

\begin{samepage}
\begin{minted}{python}
from wfl.autoparallelize import AutoparaInfo, RemoteInfo
remote_info = RemoteInfo(
    sys_name="local_HPC", 
    job_name="REF_evaluation",
    num_inputs_per_queued_job=1,
    resources={
        "num_nodes": 1, 
        "max_time": "1h",
        "partitions": "standard"},
    input_files=["pseudopotentials"])
                
autopara_info = AutoparaInfo(remote_info=remote_info)                   
\end{minted}
\end{samepage}

The \py{RemoteInfo} object specifies the system name from the \py{"$\sim$/.expyre/config.json"} file, an arbitrary name 
for labeling the submitted job, and the number of configurations from the \py{inputs} iterable to 
group for each queued job. In addition, the object specifies the required queuing system resources for each job,
one node for one hour on the partition named ``standard'' (matching the configuration file entry for this system), and 
finally, the \py{"pseudopotentials"} subdirectory will be copied for each remote job by \py{ExPyRe}.

The DFT single-point evaluation use case also has other implications for the way \py{wfl} manages
the calculations. Because \py{ASE} does most DFT calculations using 
third-party software that relies on files for input and output, \py{wfl} wraps DFT calculators such as 
\py{ase.calculators.espresso.Espresso} to make them more applicable to its intended use. It
runs each instance of the DFT program in a separate subdirectory so multiple side-by-side runs do not overwrite each other,
and can automatically select a distinct $\Gamma$-point-only executable for nonperiodic configurations.
The calculator that will be passed to \py{generic.calculate} is defined by a class inheriting from
the underlying \py{ASE} calculator, with some additional optional arguments to control the 
directories and files created during the calculation. 
\ch{Note however, that in general we do not introduce any new functionality (such as DFT convergence checks) in addition to what is provided in the underlying ASE interface and only extend the ASE's \py{Calculator}s to not interfere across parallelized instances.}
\begin{minted}{python}
from wfl.calculators.espresso import Espresso
dft_calc = Espresso(
    pseudopotentials={"Si": "Si.UPF"},
    pseudo_dir="pseudopotentials",
    input_data={
        "SYSTEM": {
            "ecutwfc": 40, 
            "input_dft": "LDA"}},
    kpts=(2, 3, 4), 
    conv_thr=0.0001)
\end{minted}
All of the arguments used here are standard \py{ase.calculators.espresso.Espresso} constructor arguments,
without any of our subclass-specific arguments that would modify the default behavior of running in a subdirectory 
named \py{"./run\_QE\_$<$RANDOM\_STR$>$"}, and keeping only files required for NOMAD\cite{Draxl_2019} upload, i.e. \py{"*.pwo"}. 

The call to do the DFT evaluations is simply
\begin{minted}{python}
from wfl.calculators import generic
evaluated_configs = generic.calculate(
    inputs, outputs, calc=dft_calc,
    autopara_info=autopara_info)
\end{minted}

This call to \py{generic.calculate} will break up the inputs into groups of one configuration each, prepare and
submit each job, and wait for them to be finished before assembling the returned configurations and
storing them in the location specified by the \py{outputs} argument.
Except for the need to wait,
\ch{executing this script with remote execution} 
should be entirely equivalent to running the same sequence of function calls locally. An optional 
\py{timeout} argument to the \py{RemoteInfo} constructor specifies the maximum waiting time. If this time is 
exceeded and an exception is raised, or the user aborts the script before all the jobs are complete, 
rerunning the \py{wfl} script will automatically resume by looking for the previously submitted jobs, 
gathering their results if they are now complete (or waiting for the rest), and continuing with any 
further actions in the script. \ch{Note that as for the reuse of output files generated by any
\py{autoparallelize} operation (Sec.~\ref{sec:using:io_abstraction}), this job caching mechanism is not aware of changes to
the code, and the user must manually wipe the staged jobs if the underlying code has been modified.}

\ch{
Given the possibility of many nested types and levels of parallelism, there can be many
ways to distribute of the work of each operation among jobs, nodes, and cores.
The number of configurations grouped into each job is specified by
\py{RemoteInfo.num\_inputs\_per\_queued\_job}, which divides the number of configurations in the input \py{ConfigSet}
to determine the number of jobs that are created and submitted to the HPC queue. 
The number of nodes allocated for each job is set by the \py{RemoteInfo.resources.num\_nodes} argument. 
Within each job, the number of operations executed in parallel is given by \py{AutoparaInfo.num\_python\_subprocesses}. 
Since controlling the full range of possible parallelism in the remote job would add a lot of complexity, we recommend two simpler configurations.
One is to use single-node jobs (i.e.\ \py{num\_nodes=1}) with \py{multiprocessing.pool}-based autoparallelization, where multiple configurations are evaluated in parallel, each using a single core (the default behavior).
The other is to use multi-node jobs with MPI-aware operations (e.g.\ DFT evaluation executables) without autoparallelizing over configurations (i.e. set \py{RemoteInfo.num\_inputs\_per\_queued\_job=1} or 
\py{AutoparaInfo.num\_python\_subprocesses=1}).
}

\subsection{Non-parallelized operations}
\label{sec:using:non_para}

Some operations that are computationally expensive, but not embarrassingly parallel have also
been wrapped in \py{wfl}, in particular the actual fitting of MLIPs. For example, fitting a GAP model can be done with
\begin{minted}{python}
from wfl.fit.gap.simple import run_gap_fit
run_gap_fit(
    fitting_configs=training_configs,
    fitting_dict=gap_fit_cli_params,
    stdout_file='gap_fit.out',
    skip_if_present=True,
    remote_info=remote_info)
\end{minted}
This function is a wrapper of the \py{gap\_fit} executable which is aware of,
but does not abstract away, its interface. The \py{gap\_fit\_cli\_params} argument is a dictionary that is converted to the command line
parameters for the \py{gap\_fit} executable.  The wrapper can detect if the GAP fit appears
to have been completed and skips the task if \py{skip\_if\_present} is set to \py{True}, but unlike autoparallelized 
operations this is implemented manually in the \py{run\_gap\_fit} function. The \py{remote\_info} argument 
specifies the queuing system resources required to run this GAP fit, e.g.\ a large memory node if
the fitting database is large, or a remote cluster if using the MPI parallel version of GAP fit. 

Since the output of this function is not a set of atomic configurations, it does
not return a \py{ConfigSet}, but only saves the resulting GAP MLIP to a file.
The name of the resulting GAP model file is specified by the standard \py{gap\_fit} 
command line argument \py{"gap\_file"}, which should be included in \py{gap\_fit\_cli\_params}.

\subsection{Daisy-chaining operations}
\label{sec:using:daisy_chain}

One important aspect of the design of \py{wfl} autoparallelized operations for workflow applications
is that they return a \py{ConfigSet}, 
so the output of each function can be used as the \py{inputs} argument of the next. For example, evaluating
the reference energy for a set of configurations and then selecting only the ones with low energy/atom can be done
with
\begin{minted}{python}
all_atoms = ConfigSet("atoms_init.xyz")

from wfl.calculators import generic
ref_eval_configs = generic.calculate(
    all_atoms,
    OutputSpec("atoms_REF_evaluated.xyz"),
    property_prefix="REF_",
    calc=dft_calc)

low_REF_E_configs = wfl.select.simple.by_bool_function(
    ref_eval_configs, 
    OutputSpec(),
    filter_func=lambda a: a.info["REF_energy"]/len(a) < 1.0)
\end{minted}

The selection call does not depend on where the output of the first call is stored, since that
information is abstracted in the \py{ref\_eval\_configs} object. 
The simple selection just assumes that \py{Atoms.info["REF\_energy"]} is defined for the configurations
in \py{ref\_eval\_configs}, and the \py{OutputSpec} specifies that
configurations returned in \py{low\_REF\_E\_configs} will be stored only in memory. Any number of
steps in the workflow can be chained similarly, and the user can have access, even after the script
is complete, to any intermediate result that was specified by a file in an operation's \py{OutputSpec} object.

\subsection{Combining elements into a workflow}
\label{sec:using:simple_workflow}

\begin{figure*}[ht]
    \centering
    \includegraphics[width=\linewidth]{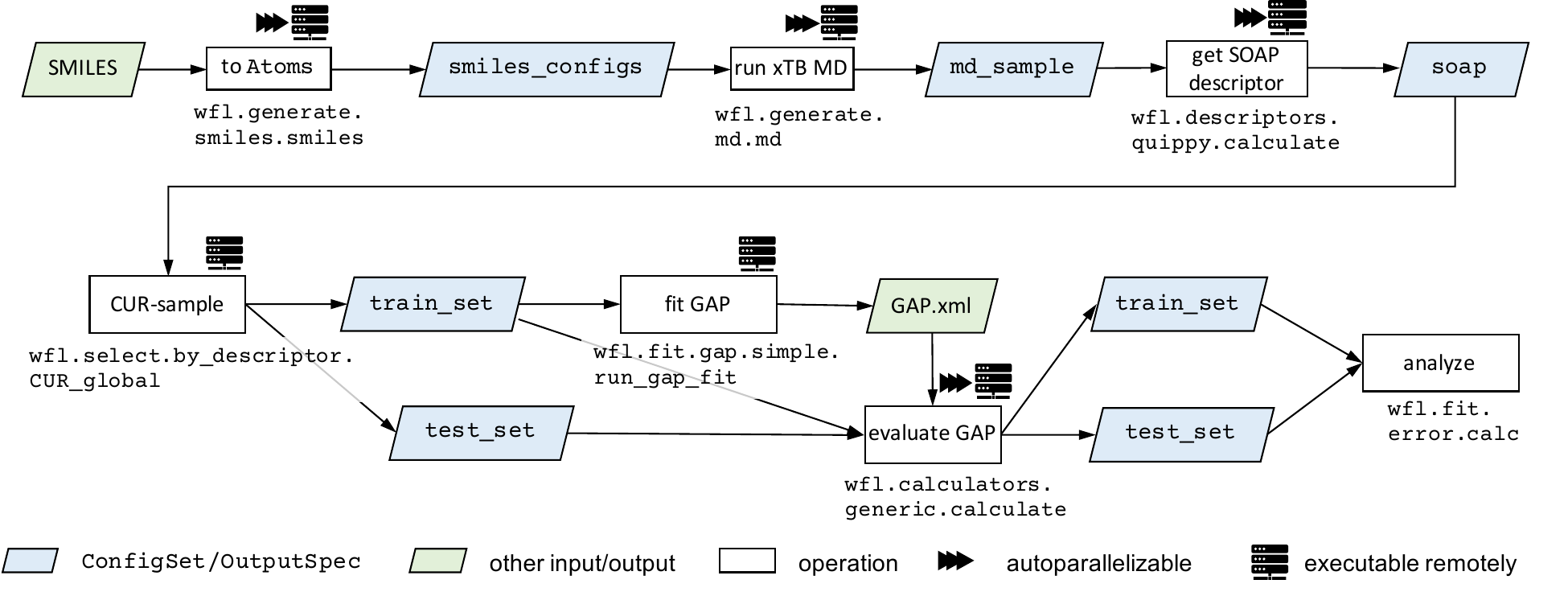}
    \caption{Schematic representation of a more complex daisy-chained workflow. The \ch{corresponding} Jupyter notebook is available online at \hbox{\url{https://libatoms.github.io/workflow/examples.daisy_chain_mlip_fitting.html}.}
    }
    \label{fig:selection_of_molecules}
\end{figure*}

An example of a more complex multi-step workflow is shown in Fig.~\ref{fig:selection_of_molecules}, and a runnable
Jupyter notebook implementing it is available in the SI and online.
After importing the needed symbols, the workflow defines
the xTB tight-binding method as the reference calculator. It creates isolated atom configurations for each
species (not shown in Fig.~\ref{fig:selection_of_molecules}),
as well as molecules defined from SMILES strings, and uses those molecules as initial configurations
for a finite temperature molecular dynamics run with the xTB calculator. The resulting trajectories are
subsampled by computing SOAP descriptors for each configuration and selecting among them with leverage score
CUR on the descriptor vectors. A fitting set is used to fit a GAP MLIP, and the predictions of the
resulting potential for the fitting set as well as an independent test set are computed.  These predictions
are then formatted into a printed error table and corresponding parity plots.

\ch{It is worth noting, however, that even though extending the new functions with \py{wfl} tools may make them more efficient and better integrate with other \py{wfl}-wrapped functions (see Section~\ref{sec:writing_functions}), any Python function may be used in a \py{wfl}-based workflow. 
For example, to make use of a new MLIP framework, only a Python function that performs the fitting and an ASE-type \py{Calculator} are needed. 
As a result, it should be straightforward to use other MLIP frameworks or packages with related functionality, such as FitSNAP~\cite{rohskopf2023fitsnap}, FLAME~\cite{amsler2020flame} and BenchML~\cite{poelking2022benchml}, within \py{wfl}-based scripts and similarly, \py{wfl}'s tools may be useful for doing operations in such frameworks.}

%% file: examples_writing.tex
\section{Wrapping new functions}
\label{sec:writing_functions}

Section \ref{sec:using_by_example} highlighted key functionality of \py{wfl} by a number of simple examples that use functions already written to take advantage of the key functionality. These included atomic structure input and output via the \py{ConfigSet} and \py{OutputSpec} classes, and controlling automatic parallelisation and executing the functions remotely. However, it is impossible, and we do not aspire, to maintain a full library of \py{wfl}-wrapped operations to support all atomistic simulation needs. Instead, we designed \py{wfl} so its tools can be easily plugged into any ASE-based script. 
Below we show examples of how to parallelize a function with \py{wfl.map} or \py{wfl.autoparallelize}. We also show how to use \py{ExPyRe} to remotely execute any function, not limited to those already in \wfl or ASE.

\subsection{Parallelizing a new function}

\subsubsection{\py{wfl.map}}
\label{sec:writing:parallelizing:map}

The simplest way to parallelize a new function over multiple \py{Atoms} objects, including taking advantage of \py{ConfigSet} and \py{OuptutSpec} and remote execution, is via \py{wfl.map.map}. 

As an example, suppose we already have a \py{cap\_bonds} function which takes a single \py{Atoms} object and adds an atom to any dangling bond: 

\begin{minted}{python}
saturated_at = cap_bonds(dangling_at, cap_with="H")
\end{minted}

\py{wfl.map.map} takes such a function, with its arguments and keyword arguments, and applies it to each structure in the input \py{ConfigSet} and writes to the output \py{OutputSpec}. 

\begin{samepage}
\begin{minted}{python}
import wfl.map
saturated_ats = wfl.map.map(
    inputs, 
    outputs, 
    map_func=cap_bonds,
    kwargs={"cap_with": "H"},
    autopara_info=autopara_info)
\end{minted}
\end{samepage}

The parallelized function has to be pickleable 
and has to take a single \py{Atoms} structure as the first positional argument and return an \py{Atoms} structure  or \py{None}. Just like examples in Section \ref{sec:using_by_example}, parallelization and remote execution of \py{wfl.map.map} is controlled by an \py{AutoparaInfo} object. 

This way of parallelizing a function is suitable in cases where starting the paralellized function is computationally inexpensive, because \py{wfl.map} calls the function separately for each \py{Atoms} object. 

\subsubsection{\py{wfl.autoparallelize}}

In some cases, the initialization step of the parallelized function is significantly more computationally expensive than executing it on a single \py{Atoms} object. For example, initializing a GAP Calculator via \py{quippy.potential.Potential} in extreme cases can take minutes (because a large parameter file has to be read in and parsed), while evaluation on a single configuration may require only a few seconds. With large numbers of atomic structures to be evaluated, it makes sense to load the potential once per tens or hundreds of structures. Furthermore, the iterable to be parallelized over might not be an atomic structure itself, but an instruction for generating one, as is the case in \py{wfl.generate.smiles} and \py{wfl.generate.buildcell} submodules. The \py{wfl.autoparallelize} wrapper supports both these scenarios.

Functions wrapped in \py{wfl.autoparallelize} have to take and return a list of \py{Atoms}, hence we define a new function
\begin{samepage}
\begin{minted}{python}
def saturate_db(input_atoms, cap_with="H"):
    output_atoms = []
    for dangling_at in input_atoms: 
        saturated_at = cap_bonds(dangling_at, 
                                 cap_with=cap_with)
        output_atoms.append(saturated_at)
    return output_atoms
\end{minted}
\end{samepage}
We can autoparallelize this function simply by calling
\begin{minted}{python}
from wfl.autoparallelize import autoparallelize
autoparallelize(saturate_db,inputs,outputs,
    cap_with="H",autopara_info=autopara_info)
\end{minted}
Or define a new, parallelized version of the function:
\begin{samepage}
\begin{minted}{python}
def autopara_saturate_db(*args, **kwargs):
    default_ap_info={"num_inputs_per_python_subprocess":100}
    return wfl.autoparallelize.autoparallelize(
             saturate_db, *args, 
             default_autopara_info=default_ap_info, 
             **kwargs)
\end{minted}
\end{samepage}

Here, before returning the wrapped function, we additionally set the default value for \py{num\_inputs\_per\_python\_subprocess}. The new \py{autopara\_saturate\_db} can then be used as in the examples in Sections \ref{sec:using_by_example} .To work with \py{autoparallelize}, the operation must take in an iterable and return a list (or nested lists) of \py{Atoms} objects or \py{None}. In most cases, iterable corresponds to a list of input \py{Atoms} objects, but can in principle be anything, for example a list of SMILES strings (e.g.\ SMILES string \py{"CCO"} corresponds to ethanol) 
used to generate  \py{Atoms} structures, as in \py{wfl.generate.smiles}.

The signature of the parallelized \py{autopara\_saturate\_db} is modified from that of \py{saturate\_db}. The iterable is replaced by  \py{inputs} (a \py{ConfigSet}) and \py{outputs} (an \py{OutputSpec}) and it accepts an \py{AutoparaInfo} object which controls parallelization. 
\begin{samepage}
\begin{minted}{python}
capped_structures = autopara_saturate_db(
    inputs, 
    outputs, 
    cap_with="H", 
    autopara_info=autopara_info)
\end{minted}
\end{samepage}

\subsection{Execute Python Remotely}
\label{sec:writing:expyre}

\begin{figure}[ht]
    \centering
    \includegraphics[width=0.8\linewidth]{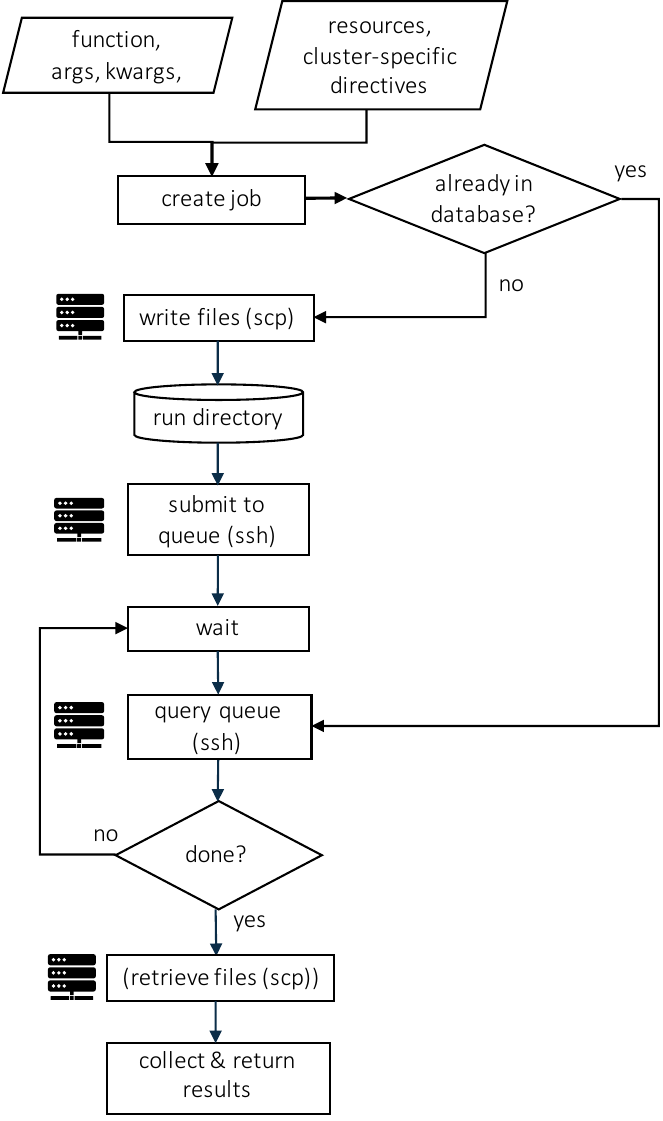}
    \caption{Illustration of steps performed by \py{ExPyRe} when executing functions remotely. \ch{Pictogram indicates} actions \ch{that} may be executed on a remote cluster. \markuponly{Revision markup only: Note the change in the diagram ``hash'' to ``create job'' and change of color-coding to pictogram.}}
    \label{fig:expyre}
\end{figure}

Running remote jobs in \wfl is done using \py{ExPyRe}, which wraps individual Python functions, executes them in remote jobs,
and returns their results.  It is inspired by MyQueue\cite{mortensen2020myqueue}, but designed with somewhat different
limitations and capabilities motivated by our functionality and computational resource use cases.
The autoparallelizing wrapper ensures that all autoparallelized functions interface with \py{ExPyRe}.
In addition, some non-\py{autoparallelize}-able yet computationally expensive tasks, such as fitting GAP and ACE interatomic potentials, are an important part of atomistic workflows and are also interfaced with \py{ExPyRe} directly. The key steps performed by \py{ExPyRe} are illustrated in Figure \ref{fig:expyre}.

To execute any function with \py{ExPyRe} (let us continue with the \py{cap\_bonds} example), a couple of extra Python objects need to be defined, as illustrated in the following script. Note that \py{RemoteInfo} is not applicable here, because it combines \py{wfl}-specific options with \py{ExPyRe} arguments.

\begin{samepage}
\begin{minted}{python}
xpr = expyre.func.ExPyRe('test_task', 
             function=cap_bonds, 
             args=[dangling_at], 
             kwargs={'cap_with': "H"})

resources = {'max_time': '1h', 
             'num_nodes': 1, 
             'partitions': 'regular'}
       
xpr.start(resources=resources, system_name='remote')
saturated_at, stdout, stderr = xpr.get_results()
\end{minted}
\end{samepage}

First, an \py{ExPyRe} instance is created in which we specify the function that will be executed remotely, as well as its positional and keyword arguments. Together with \py{xpr.start} these steps write input files for the job, write a queue-specific job submission script and submit the job.
In this step, job-specific resources (required time, number of nodes or cores, queue or partition) are also provided and used to specify the correct resources in the job submission script. As in the examples from Section \ref{sec:using_by_example}, the cluster-specific configuration file must be present.  

At the \py{xpr.get\_results} stage, \py{ExPyRe} periodically queries the queueing system for status of the job while waiting for its completion. At this point, the Python script may be killed and restarted to continue without re-submitting the jobs. 
The jobs and their status are tracked in a file-based sqlite database which can be queried via a convenience command line interface, \py{xpr}. 

File copying and job queue querying is executed via a passwordless ssh, for example by using public/private keys, kerberos authentication, or by multiplexing SSH connections to reuse a previously established password- or multifactor-authenticated connection. 

There are several files associated with the remote execution process used for communication between the local Python script and remote queued job. 
These files include the pickled input function and its output, the job submission script and outputs as well as files used by \py{ExPyRe} to track and report on job's progress.
Normally, the user does not need to be concerned with these files, but
it is useful to be aware of the behind-the-scenes working of \py{ExPyRe} for occasional debugging, since we expect these tools to be used in developing a wide variety of projects and scripts. These auxiliary files can also be deleted using the command line interface, \py{xpr}.

\ch{
One consequence of the \py{pickle}-based remote execution mechanism is that the function to be executed remotely 
(e.g.\ \py{cap\_bonds} or \py{saturate\_db}) must be {\em imported} into the script that wraps it with 
\py{wfl.autoparallelize} or \py{wfl.map} (or any other way that results into passing it to \py{ExPyRe}).
%
%
This is because pickling a function does not save the function's content, only its full name including package
information, which is used to {\em import} it when the pickled data is unpickled by the remotely executed
job.
%
%
The solution is to define the original function either in a Python package that is installed on both the local and remote machines, or in a separate file that is included in \py{RemoteInfo.input\_files} and therefore copied over to the run directory on the remote machine. 
}

%% file: outro.tex
\section{Summary}
\label{sec:summary and outlook}

In summary, we introduced the Python-based workflow management package \wfl and the remote execution package \py{ExPyRe}, which are designed to support the versatile requirements of computations commonly used in atomistic simulation and MLIP fitting processes. 
\wfl is designed as a lightweight platform to provide efficient parallelization capabilities for various numbers and sizes of embarrassingly parallel tasks with computational demands that range from milliseconds on a single core up to hours on several nodes. These parallel tasks can optionally be executed on a HPC queueing system via the general-purpose Python remote execution framework \py{ExPyRe}. The two packages provide a few low-level functions - input and output abstraction classes, an autoparallelization wrapper, and a remote execution functionality - to wrap any \emph{operation}, as well as wrapped versions of a number of
commonly used operations. 
Using this developer-oriented functionality, high level user-friendly functions can be easily produced by combining these
and other operations to construct reproducible calculation workflows. Some common instances of such high level functions are already present in the \wfl package, and we provide examples of how to make use of \py{wfl}'s functionality for custom functions and workflows.

In contrast to other material simulation workflow packages that focus on \emph{ab initio} data generation and provenance,
\wfl is designed to provide low-level support for efficiently parallelizing and integrating  a wide variety of operations for atomistic simulations.
One distinctive feature is its focus on using human-readable formats and storage, and on minimizing the need
for system infrastructure such as remotely accessible database servers.
Another is the
use of a small number of python classes to abstract atomic configuration
storage and to support efficient execution of embarrassingly parallel operations, whether they are computationally
demanding individually or only in the aggregate. 
The package is developer-oriented, and intended for use in research areas where the development of the particular sequence of computational operations to be done is a complex and significant part of the overall research task. 

We believe that \wfl will fill an important niche in the atomistic simulation communities that develop new computational procedures
with tools that implement them using Python libraries such as ASE. While the basic functionality of ASE has already revolutionized
the way atomistic simulation software is being developed, \wfl will extend its range of applicability further by abstracting
the details of atomic configuration, and providing easy to use automatic parallelization and remote execution. Its design
principles of lightweight, modularity, and human-inspectability should make it easy to incorporate into existing ASE-based
scripts. We hope that it will prove useful for a wide range of atomistic simulations.

\section{Code Availability}

    The \wfl and \py{ExPyRe} codes are available in the GitHub respositories, \url{https://github.com/libAtoms/workflow} and \url{https://github.com/libAtoms/ExPyRe}. Additionally, documentation, including examples and API, are accessable at \url{https://libatoms.github.io/workflow} and \url{https://libatoms.github.io/ExPyRe}.

\section{Data Availability}

    Data sharing is not applicable to this article as no new data were created or analyzed in this study. 

\section{Acknowledgements}

We are grateful for computational support from the UK national high-performance computing service, ARCHER2, for which access was obtained via the UKCP consortium and funded by EPSRC grant reference EP/P022596/1. 
N.B. acknowledges support by the U.~S. Naval Research Laboratory's 6.1 fundamental research base program, and computer time support from the DOD HPCMPO at the AFRL, ERDC, and ARL MSRCs. 
E.G. acknowledges support from the EPSRC Centre for Doctoral Training in Automated Chemical Synthesis Enabled by Digital Molecular Technologies with grant reference EP/S024220/1.
T.K.S.\ acknowledges support from the European Union’s Horizon 2020 Research and Innovation Program under Grant Agreement No.\ 957189 (BIG-MAP project).
S.W., H.H. and K.R. gratefully acknowledge the Max Planck Computing and Data Facility (MPCDF) for providing computing time. 

For their contributions to the code development the authors want to acknowledge Nikhil Bapat, Lars Schaaf, Nicolas Bergmann, Xu Han and Felix Riccius. We also thank Olga Vinogradova
and Felix Riccius for creating the logo.

For the purpose of open access, the authors have applied a Creative Commons Attribution (CC BY) licence to any Author Accepted Manuscript version arising from this submission.

\section{Author Contribution}

    G.C., N.B. and T.K.S. jointly conceived the project. N.B., E.G., S.W. and T.K.S. designed, implemented and tested the software. N.B., G.C., H.H.H. and K.R. supervised the project, N.B., G.C. and K.R. acquired funding. E.G., N.B. and H.H.H. prepared the original draft of the paper. All authors reviewed and edited the manuscript.

\section{Competing interests}

    Authors declare no competing interests.